\documentstyle[12pt]{article}
\input epsf
\begin{document}

\newcommand{\vp}{\varphi}
\newcommand{\be}{\begin{equation}} 
\newcommand{\ee}{\end{equation}}
\begin{center}
{\bf TOPOLOGICAL DEFECTS AND THE FORMATION OF STRUCTURE IN THE UNIVERSE}\footnote[1]{Invited lecture at the {\it Pacific Conference on Gravitation and Cosmology}, Feb. 1 - 6 1996, Seoul, Korea, to be publ. in the proceedings (World Scientific, Singapore, 1996).}
 
\vskip 1cm

Robert BRANDENBERGER\footnote[2]{rhb@het.brown.edu.} 

\vskip 1cm

{\it Physics Department, Brown University\\
 Providence, RI. 02912, USA.}

\end{center}
\vskip 1cm

\noindent{\bf Abstract}
\vskip 0.7cm
Topological defects, in particular cosmic strings, give rise to an interesting mechanism for generating the primordial perturbations in the early Universe which are required to explain the present structure.
An overview of the cosmic string model will be given, focusing on the predictions of the theory for the large-scale structure of the Universe and for cosmic microwave background anisotropies.

\vskip 1cm

\noindent{\bf I.  Introduction}
\vskip 0.7cm
Over the past fifteen years, we have witnessed the development of the first
theories based on causal microphysics which attempt to explain the observed
structure in the Universe and make predictions for observations on even larger
scales than those for which we now have information \cite{review}. The class of models which is receiving most attention is 
based on inflation. According to the inflationary Universe scenario,
quantum fluctuations during a period of exponential expansion in the very early Universe generate a random phase and approximately scale-invariant spectrum of density fluctuations on scales which are cosmological today.

Inflation, however, is not without its own problems, and thus it is of great interest to have an alternate class of theories based on physics more accessible to experiment. Topological defect models provide such a class of models. In defect models, topological defects which form during a phase transition in the very early Universe provide the seeds about which galaxies and even larger structures form by gravitational clustering. The most promising of the defect models is the cosmic string theory \cite{VSbook}.

In this lecture I will give an overview of the cosmic string theory of structure formation, focusing on the predictions of the model for large-scale structure and cosmic microwave background (CMB) anisotropy observations. For detailed expositions of the theory, the reader is refered to several recent 
publications \cite{VSbook}, \cite{review}.

Any cosmological model requires (at least) two inputs, namely both the specification of the source of perturbations and of the dark matter which makes up the bulk of the matter in the Universe. In inflation-based theories, most of the dark matter must be cold (i.e. the corresponding particles are nonrelativistic at the time of equal matter and radiation $t_{eq}$, the time when density fluctuations begin to increase in amplitude) whereas in the cosmic string theory it is possible (as will be discussed below) for the dark matter to be hot (i.e. the corresponding particles are relativistic at $t_{eq}$).
\vskip 1cm
\noindent{{\bf II.  Defects and their Classification}
\vskip 0.7cm
The topological defects of interest in cosmology are those which arise in relativistic field theories. Due to the high temperature and density of the early Universe, relativistic field theory is believed to give the correct description of matter. Based on particle physics experiments it is also believed that the Universe underwent a series of phase transitions in the early Universe during which internal symmetries of matter fields were spontaneously broken. This symmetry breaking is described by an order parameter $\varphi$ (which does not necessarily have to be a scalar field but which - as in condensed matter physics - could represent a condensate of fermionic fields) with a potential of the
form 
\begin{equation}
\label{one}
V (\varphi) = {1\over 4} \lambda (\varphi^2 - \eta^2)^2  
\end{equation}
The phase transition will take place on a short time
scale $\tau < H^{-1}$, and will lead to correlation regions of radius $\xi <
t$ inside of which $\varphi$ is approximately constant, but outside of which
$\varphi$ ranges randomly over the vacuum manifold ${\cal M}$, the set of
values
of $\varphi$ which minimize $V(\varphi)$ -- in our example $|\varphi|
= \eta$.  The correlation regions are separated by regions in
space where $\varphi$ leaves the vacuum manifold ${\cal M}$ and where,
therefore, potential energy is localized.  Via the usual gravitational
force, this energy density can act as a
seed for structure.

Topological defects are familiar from solid state and condensed matter
systems.  Crystal defects, for example, form when water freezes or
when a metal crystallizes.  Point defects, line defects and planar
defects are possible.  Defects are also common in liquid crystals.
They arise in a temperature quench from the disordered to the ordered
phase.  Vortices in $^4$He are analogs of global cosmic strings.
Vortices and other defects are also produced during a quench below the
critical temperature in $^3$He.  Finally, vortex lines also play an
important role in the theory of superconductivity.

The analogies between defects in particle physics and condensed matter
physics are quite deep.  Defects form for the same reason: the vacuum
manifold is topologically nontrivial.  The arguments which say that in
a theory which admits defects, such defects will inevitably form \cite{Kibble}, are
applicable both in cosmology and in condensed matter physics.
Different, however, is the defect dynamics.  The motion of defects in
condensed matter systems is friction-dominated, whereas the defects in
cosmology obey relativistic equations, second order in time, since they come from a relativistic field theory.

To classify the possible topological defects, we consider theories with an $n$-component order
parameter $\varphi$ and with a potential energy function (free energy
density) of the form (\ref{one}) with
$\varphi^2$ denoting the sum of the squares of all of the components.
There are various types of topological defects
(regions of trapped energy density) depending on the number $n$ of components
of $\varphi$ \cite{Kibble}. For a single component real scalar field as order parameter (i.e. $n = 1$), the defects are domain walls, for $n = 2$ the defect configurations are one dimensional strings (cosmic strings), and for $n = 3$ monopoles result.
In theories with a global symmetry and $n = 4$ it is possible to have textures, defects in space-time (collapsing lumps of scalar field gradient energy in space).

Theories with domain walls forming above the scale of electroweak symmetry breaking are ruled out since a single domain wall
stretching
across the present horizon would overclose the Universe.  Local monopoles corresponding to a scale of symmetry breaking comparable to the scale of grand unification are
also ruled out since they would likewise overclose the Universe. Promising theories from the point of view of cosmology are models with cosmic strings, global monopoles or textures (for a review of the latter see \cite{Turok}). In the following, I will focus on cosmic strings.

A theory with a complex order parameter $(n = 2)$ admits
cosmic strings.  In this case the vacuum manifold of the
model is ${\cal M} = S^1$,
which has nonvanishing first homotopy group $\Pi_1 ({\cal M}) = Z \neq 1$.
A cosmic string is a line of trapped energy density which arises
whenever the field $\varphi (x)$ circles ${\cal M}$ along a closed path
in space ({\it e.g.}, along a circle). To construct a field configuration with a string along the $z$ axis,
take $\varphi (x)$ to cover ${\cal M}$ along a circle with radius $r$
about the point $(x,y) = (0,0)$:
\be
\label{two}
\varphi (r, \vartheta ) \simeq \eta e^{i \vartheta} \, , \, r \gg
\eta^{-1} \, .
\ee
This configuration has winding number 1, {\it i.e.}, it covers ${\cal
M}$ exactly once.  Maintaining cylindrical symmetry, we can extend
(\ref{two}) to arbitrary $r$
\be
\label{three}
\varphi (r, \, \vartheta) = f (r) e^{i \vartheta} \, , 
\ee
where $f(0) = 0$ and $f(r)$ tends to $\eta$ for large $r$.  The
width $w$ of the string is defined as the value of $r$ beyond which 
$\eta - f(r) < {1 \over 2} \eta$. It can be estimated by balancing potential and tension energy.  The result is 
\be
\label{four}
w \sim \lambda^{-1/2} \eta^{-1}.
\ee
For local cosmic strings, {\it i.e.}, strings arising due to the
spontaneous breaking of a gauge symmetry, the energy density decays
exponentially for $r \gg w$.  In this case, the energy $\mu$
per unit length of a string is finite and depends only on the symmetry
breaking scale $\eta$
\be
\label{five}
\mu \sim \eta^2 
\ee
(independent of the coupling $\lambda$).  The value of $\mu$ is the
only free parameter in a cosmic string model.
\vskip 1cm
  
\noindent{\bf III.  Formation of Topological Defects}
\vskip 0.7cm
The Kibble mechanism \cite{Kibble} ensures that in theories which admit
topological defects, such defects will inevitably be produced
during a phase transition in the very early Universe.

Consider a mechanical toy model (see Fig. 1), first introduced by Mazenko, Unruh
and Wald \cite{MTW} in the context of inflationary Universe models, which
is useful in understanding the scalar field evolution.  Consider a lattice of points on a flat table .  At each point, a pencil
is pivoted.  It is free to rotate and oscillate.  The tips of nearest
neighbor pencils are connected with springs (to mimic the spatial
gradient terms in the scalar field Lagrangean).  Newtonian gravity
creates a potential energy $V(\varphi)$ for each pencil ($\varphi$ is
the angle relative to the vertical direction).  $V(\varphi)$ is
minimized for $| \varphi | = \eta$ (in our toy model $\eta = \pi /
2$).  Hence, the Lagrangean of this pencil model is analogous to that
of a scalar field with symmetry breaking potential (\ref{one}).

\smallskip \epsfxsize=8cm \epsfbox{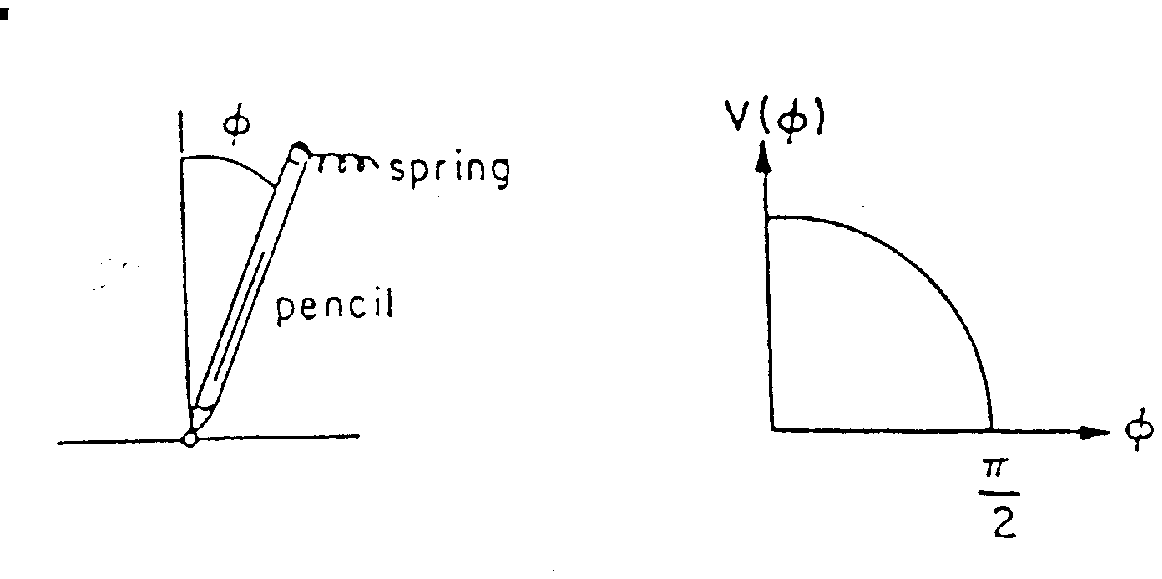}
{\baselineskip=13pt
\noindent{\bf Figure 1:} The pencil model: the potential energy of a
single pencil has the same form as that of scalar fields used for
spontaneous symmetry breaking.  The springs connecting nearest
neighbor pencils give rise to contributions to the energy which mimic
spatial gradient terms in field theory.}

\medskip
At high temperatures $T \gg T_c$, all pencils undergo large amplitude,
high frequency oscillations.  However, by causality, the phases of
oscillation of pencils with large separation $s$ are uncorrelated.
For a system in thermal equilibrium, the length $s$ beyond which
phases are random is the correlation length $\xi (t)$.  However, since
the system is quenched rapidly, there is a causality bound on
$\xi$:
\be
\xi (t) < t \, ,  
\ee
where $t$ is the causal horizon.

The critical temperature $T_c$ is the temperature at which the
free energy of a pencil is equal in the vertical and horizontal positions.  
For $T < T_c$, it is energetically preferable for a pencil to
lie flat on the table.  However, the orientations of the pencils are random beyond
a distance of $\xi (t)$ determined by equating the free energy gained by
symmetry breaking (a volume effect) with the gradient energy lost (a surface
effect).  As expected, $\xi (T)$ diverges at $T_c$. Very close to $T_c$, the
thermal energy $T$ is larger than the volume energy gain $E_{corr}$ in a
correlation volume. Hence, these domains are unstable to thermal fluctuations.
As $T$ decreases, the thermal energy decreases more rapidly than $E_{corr}$.
Below the Ginsburg temperature $T_G$, there
is insufficient thermal energy to excite a correlation volume into the
state $\varphi = 0$.  Domains of size
$\xi (t_G) \sim \lambda^{-1} \eta^{-1}$
freeze out \cite{Kibble},\cite{Kibble2}.  The boundaries between these domains become
topological defects. An improved version of this argument has recently been given by Zurek \cite{Zurek} (see also \cite{BD}).

We conclude that in a theory in which a symmetry breaking phase
transition satisfies the topological criteria for the existence of a
fixed type of defect, a network of such defects will form during the
phase transition and will freeze out at the Ginsburg temperature.  The causality bound implies that the initial correlation length obeys
$\xi (t_G) < t_G$.
For times $t > t_G$ the evolution of the network of defects may be
complicated (as for cosmic strings) or trivial (as for textures).  In
any case , the causality bound
persists at late times and states that even at late times, the mean
separation and length scale of defects is bounded by $\xi (t) \leq t$.

Applied to cosmic strings, the Kibble mechanism implies that at the
time of the phase transition, a network of cosmic strings with typical
step length $\xi (t_G)$ will form.  According to numerical
simulations \cite{VV}, about 80\% of the initial energy is in infinite
strings (strings with curvature radius larger than the Hubble radius) and 20\% in closed loops.

The evolution of the cosmic string network for $t > t_G$ is
complicated .  The key processes are loop production
by intersections of infinite strings and loop shrinking
by gravitational radiation.  These two processes combine to create a
mechanism by which the infinite string network loses energy (and
length as measured in comoving coordinates). As
a consequence, the correlation length of the string network is always
proportional to its causality limit
\be
\xi (t) \sim t \, . 
\ee
Hence, the energy density $\rho_\infty (t)$ in long strings is a fixed
fraction of the background energy density $\rho_c (t)$
\be
{\rho_\infty (t)\over{\rho_c (t)}} \sim G \mu \, .  
\ee

We conclude that the cosmic string network approaches a ``scaling
solution" in which the statistical properties of the
network are time independent if all distances are scaled to the
horizon distance.
Although the qualitative characteristics of the cosmic string scaling
solution are well established, the quantitative details are not.  The
main reason for this is the fact that the Nambu action, the action
which describes the evolution of cosmic strings, breaks down at
kinks and cusps.  However, kinks and cusps inevitably form and are
responsible for the small-scale structure on strings. Hence, neither the
exact number of segments of the long string network which on average cross
a Hubble volume, nor the
amount of small-scale structure on strings is known. This is at the
present time the main obstacle to developing the cosmic string theory in
greater quantitative detail.
\vskip 1cm
\noindent{\bf IV.  Cosmic Strings and Structure Formation}
\vskip 0.7cm
The starting point of the structure formation scenario in the cosmic
string theory is the scaling solution for the cosmic string network,
according to which at all times $t$ (in particular at $t_{eq}$, the
time when perturbations can start to grow) there will be a few long
strings crossing each Hubble volume, plus a distribution of loops of
radius $R \ll t$ (see Fig. 2).

\smallskip \epsfxsize=6.5cm \epsfbox{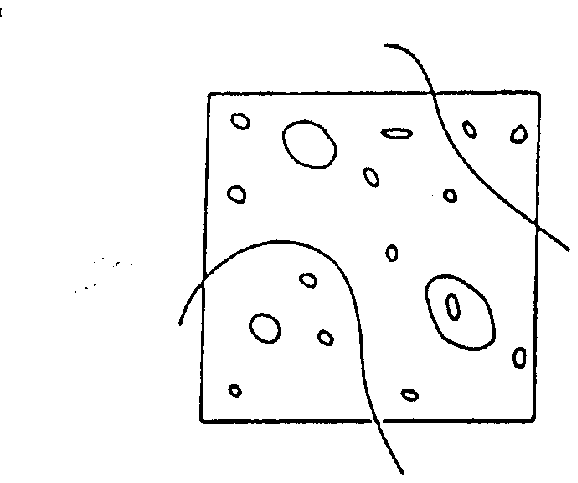}
{\baselineskip=13pt
\noindent {\bf Figure 2.}  Sketch of the scaling solution for the
cosmic string network.  The box corresponds to one Hubble volume at
arbitrary time $t$.}
\medskip 

The cosmic string model admits three mechanisms for structure
formation:  loops, filaments, and wakes.  Cosmic string loops have the same
time averaged field as a point source with mass. Hence, loops will be seeds
for spherical accretion of dust and radiation. However, according to the
recent cosmic string evolution simulations, most of the mass in strings
is in the long string network, and hence the loop mechanism is a subdominant mechanism of structure formation.
 
\smallskip \epsfxsize=9.5cm \epsfbox{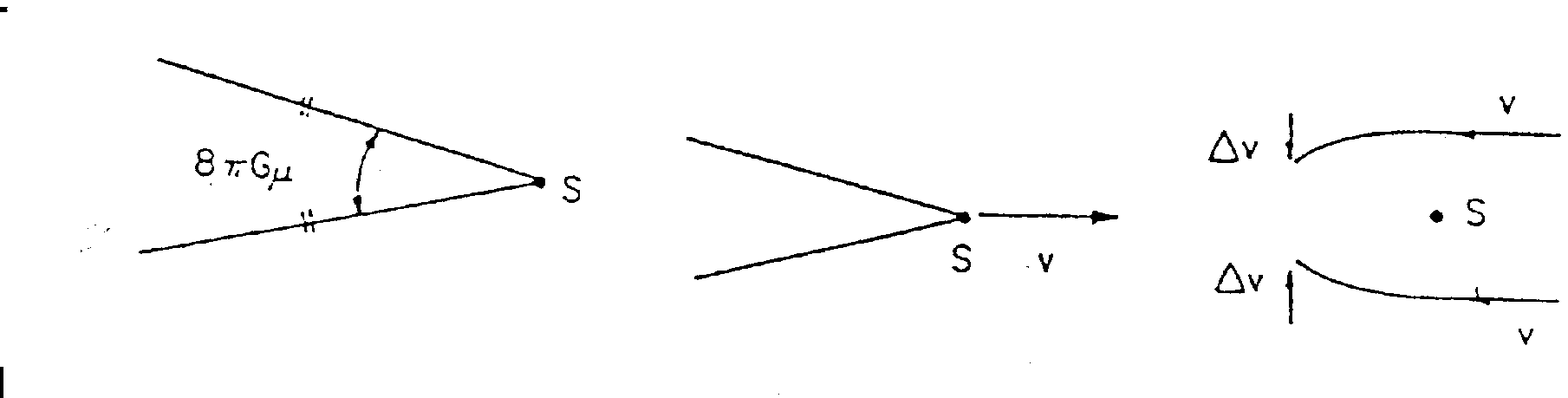}
{\baselineskip=13pt
\noindent{\bf Figure 3.} Sketch of the mechanism by which a long
straight cosmic string $S$ moving with velocity $v$ in transverse
direction through a plasma induces a velocity perturbation $\Delta v$
towards the wake. Shown on the left is the deficit angle, in the
center is a sketch of the string moving in the plasma, and on the
right is the sketch of how the plasma moves in the frame in which the
string is at rest.}  
\medskip

The second mechanism involves long strings moving with relativistic
speed in their normal plane, giving rise to
velocity perturbations in their wake \cite{SV}.  The mechanism is illustrated in
Fig. 3:
space normal to the string is a cone with deficit angle \cite{deficit}
\be
\label{angle}
\alpha = 8 \pi G \mu \, .  
\ee
If the string is moving with normal velocity $v$ through a bath of dark
matter, a velocity perturbation
\be
\delta v = 4 \pi G \mu v \gamma  
\ee
[with $\gamma = (1 - v^2)^{-1/2}$] towards the plane behind the string
results.  At times after $t_{eq}$, this induces planar overdensities,
the most
prominent ({\it i.e.}, thickest at the present time) and numerous of which were
created at $t_{eq}$, the time of equal matter and
radiation \cite{TV},\cite{SVBST},\cite{PBS}.  The
corresponding planar dimensions are (in comoving coordinates)
\be
t_{eq} z (t_{eq}) \times t_{eq} z (t_{eq}) v \sim (40 \times 40 v) \,
{\rm Mpc}^2 \, .  
\ee

The thickness $d$ of these wakes can be calculated using the
Zel'dovich approximation \cite{Zeldovich}.  The result is \cite{PBS}
\be
d \simeq G \mu v \gamma (v) z (t_{eq})^2 \, t_{eq} \simeq 4 v \, {\rm
Mpc} \, .  
\ee
\par
Wakes arise if there is little small scale structure on the string.
In this case, the string tension equals the mass density, the string
moves at relativistic speeds, and there is no local gravitational
attraction towards the string.

In contrast, if there is small scale structure on strings,
then the string tension $T$ is smaller \cite{Carter} than the mass per unit
length $\mu$ and there
is a gravitational force towards the string which gives rise to
cylindrical accretion, thus producing filaments.

Which of the mechanisms -- filaments or wakes -- dominates is
determined by the competition between the velocity induced by the gravitational potential on the string
and the velocity perturbation of the wake.   

By the same argument as for wakes, the most numerous and prominent
filaments will have the distinguished scale
\be
t_{eq} z (t_{eq}) \times d_f \times d_f  
\ee
where $d_f$ can be calculated using the Zel'dovich approximation \cite{Aguirre}.

The cosmic string model predicts a scale-invariant spectrum of density
perturbations, exactly like inflationary Universe models but for a
rather different reason.  Consider the {\it r.m.s.} mass fluctuations
on a length scale $2 \pi k^{-1}$ at the time $t_H (k)$ when this scale
enters the Hubble radius.  From the cosmic string scaling solution it
follows that a fixed ({\it i.e.}, $t_H (k)$ independent) number
$\tilde v$ of strings of length of the order $t_H (k)$ contribute to
the mass excess $\delta M (k, \, t_H (k))$.  Thus
\be
{\delta M\over M} \, (k, \, t_H (k)) \sim \, {\tilde v \mu t_H
(k)\over{G^{-1} t^{-2}_H (k) t^3_H (k)}} \sim \tilde v \, G \mu \, .
\ee
Note that the above argument predicting a scale invariant spectrum
will hold for all topological defect models which have a scaling
solution, in particular also for global monopoles and textures.

The amplitude of the {\it r.m.s.} mass fluctuations (equivalently: of
the power spectrum) can be used to normalize $G \mu$.  Since today on
galaxy cluster scales ${\delta M\over M} (k, \, t_0) \sim 1 $,
the linear theory for the growth of fluctuations yields \cite{ZelVil},\cite{TB}
\be
{\delta M\over M} \, (k, \, t_{eq}) \sim 10^{-4} \, , 
\ee
and therefore, using $\tilde v \sim 10$,
\be
G \mu \sim 10^{-5} \, .  
\ee
Thus, if cosmic strings are to be relevant for structure formation,
they must arise due to a symmetry breaking at energy scale $\eta
\simeq 10^{16}$GeV.  This scale happens to be the scale of unification (GUT)
of weak, strong and electromagnetic interactions.  It is tantalizing
to speculate that cosmology is telling us that there indeed was new
physics at the GUT scale.

A big difference of the cosmic string model compared to inflation-based theories
is that HDM is a viable dark matter candidate.  The neutrino free streaming length decreases as $a(t)^{- 1/2}$. Whereas in inflation-based HDM models, all perturbations on scales smaller than the maximal neutrino free streaming scale are erased before $t_{eq}$, cosmic string seeds survive and can seed
structures on small scales. Growth of perturbations on small scales is delayed (compared to models with CDM) but not prevented.  Accretion of hot dark matter by string wakes
was studied in Ref. \cite{PBS}. In this case, nonlinear perturbations
develop only late.  Accretion onto loops and small scale structure on the long strings provide two mechanisms which may lead to high redshift objects such as quasars and high redshift galaxies. The first mechanism has recently been studied in Ref. \cite{MB}.

The power spectra in the cosmic string models with CDM and HDM are
obviously different on scales smaller than the maximal neutrino free
streaming length. The power spectrum in a model with cosmic strings
and HDM has less power on small scales than an inflationary CDM model. Recent numerical simulations \cite{AS},\cite{Mahonen} 
demonstrate show that a COBE normalized cosmic string theory with HDM has a power spectrum in better agreement with the recent APM observational results than a standard inflationary CDM model.
\vskip 1cm
\noindent{\bf V.  Specific Signatures}
\vskip 0.7cm
Topological defect models give rise to some specific signatures and
are hence falsifiable - a condition on any good scientific theory.
The cleanest specific signatures can be found in the microwave background, 
although the signatures in the predicted large-scale structure are perhaps 
more easily detectable.

All theories of structure formation give rise to Sachs-Wolfe type
CMB temperature fluctuations. In the cosmic string model there are, in addition, specific signatures which cannot
be described in a linear perturbative analysis.
As described in the previous section, space perpendicular to a long straight
cosmic string is conical with deficit angle given by (\ref{angle}).  Consider
now CMB radiation approaching an observer in a direction normal to the
plane spanned by the string and its velocity vector (see Fig. 4).
Photons arriving at the observer having passed on different sides of
the string will obtain a relative Doppler shift which translates into
a temperature discontinuity of amplitude \cite{KS}
\be
{\delta T\over T} = 4 \pi G \mu v \gamma (v) \, ,  
\ee
where $v$ is the velocity of the string.  Thus, the distinctive
signature for cosmic strings in the microwave sky are line
discontinuities in $T$ of the above magnitude.

\smallskip \epsfxsize=7cm \epsfbox{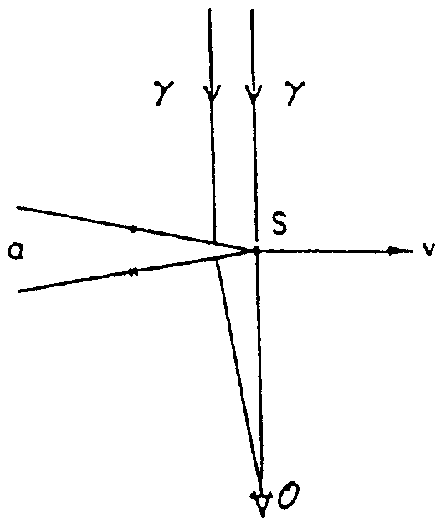}
{\baselineskip=13pt
\noindent{\bf Figure 4:} Sketch of the Kaiser-Stebbins effect by
which cosmic strings produce linear discontinuities in the CMB. Photons
$\gamma$ passing on different sides of a moving string $S$ (velocity $v$)
towards the observer ${\cal O}$ receive a relative Doppler shift due to the
conical nature of space perpendicular to the string (deficit angle $\alpha$).}
\medskip  

Given ideal maps of the CMB sky it would be easy to detect strings.
However, real experiments have finite beam width.  Taking into account
averaging over a scale corresponding to the beam width will smear out
the discontinuity, and it turns out to be surprisingly hard to
distinguish the predictions of the cosmic string model from that of
inflation-based theories using quantitative statistics which are easy
to evaluate analytically, such as the kurtosis of the spatial gradient
map of the CMB \cite{MPB}.

Other topological defect models predict different distinctive signatures.
Textures, for example, produce a distribution of hot and cold spots on the CMB sky
with typical size of several degrees \cite{TS}.  This signature is much easier
to see in CMB maps. Inflationary models typically predict a Gaussian random
field distribution of temperature fluctuations with a scale invariant spectrum.

As discussed in the previous section, the distinctive prediction of the cosmic
string theory for large-scale structure is that on scales larger than the
comoving Hubble radius at $t_{eq}$, the distribution of galaxies should be
dominated by a network of sheets. The regions between the sheets form voids
which should also be quite empty of dwarf galaxies. These predictions can
be quantified using topological statistics. In contrast, in inflation-based models, the topology of the large-scale structure should show no difference on
the above scales between overdense and underdense regions.
\vskip 1cm
\noindent{\bf VI. Conclusions}
\vskip 0.7cm
The cosmic string theory of structure formation appears at the present time
to be a viable alternative to inflation-based models. Note however, that this
theory (in contrast to inflation) does not address other cosmological problems
such as the flatness and horizon problems.

The cosmic string model predicts a scale-invariant primordial spectrum of density fluctuations. With hot dark matter, the power on scales smaller than
the neutrino free streaming length at $t_{eq}$ is reduced, but not totally
wiped out. In principle, the theory contains only one free parameter, $G \mu$.
The normalizations of $G \mu$ from the COBE measurement of CMB anisotropies and from the power spectrum on cluster scales is consistent according to the present calculations \cite{Periv} (which should be further refined). Recent 
\cite{Doppler} work indicates that
the position of the ``Doppler peaks" of the CMB power spectrum might allow
a distinction between inflation-based and topological defect models. However, more work needs to be done to improve the calculations.

There are several distinctive predictions of the cosmic string theory:
line discontinuities in the CMB temperature maps and planar topology of
the mass distribution on large scales, to name just two. Recent observations
\cite{LSS} of structure of the Universe on scales of greater than $100 h^{-1}$Mpc
are showing some evidence that the distribution of galaxies on large scales is
indeed planar. This gives further motivation to tackle the difficult problem
of making the predictions of the model more quantitative.

\centerline{\bf Acknowledgements}

I am grateful to the organizers of the {\it Pacific Conference on Gravitation
and Cosmology} for the invitation to speak at the meeting and for their
warm hospitality. This research on which this report is based has been
supported in part by the U.S. Department of Energy under Grant
DE-FG0291ER40688, Task A.

\end{document}